\documentclass[submission, Phys]{SciPost}

\begin{document}

\begin{center}{\Large \textbf{
First indication on self-similarity of strangeness production \\
in $Au+Au$ collisions at RHIC: \\
Search for signature of phase transition in nuclear matter\
}}\end{center}

\begin{center}
M. V. Tokarev\textsuperscript{1,2*},
I. Zborovsk\'{y}\textsuperscript{3**}
\end{center}

\begin{center}
{\bf 1} Joint Institute for Nuclear Research, Dubna, 141980, Russia
\\
{\bf 2} Dubna State University, Dubna, 141980, Russia
\\
{\bf 3} The Czech Academy of Science,  Nuclear Physics Institute, \\
\v {R}e\v {z}, 250 68, Czech Republic
\\
* tokarev@jinr.ru \\
** zborovsky@ujf.cas.cz
\end{center}

\begin{center}
\today
\end{center}



\definecolor{palegray}{gray}{0.95}
\begin{center}
\colorbox{palegray}{
  \begin{tabular}{rr}
  \begin{minipage}{0.1\textwidth}
  \end{minipage}
  &
  \begin{minipage}{0.75\textwidth}
    \begin{center}
    {\it 50th International Symposium on Multiparticle Dynamics}\\ {\it (ISMD2021)}\\
    {\it 12-16 July 2021} \\
    \doi{10.21468/SciPostPhysProc.?}\\
    \end{center}
  \end{minipage}
\end{tabular}
}
\end{center}

\section*{Abstract}
{\bf
New results of analysis of $K^0_S-$meson spectra measured over a wide range of energy 
$\sqrt {s_{NN}}=7.7-200$ GeV  and centrality in $Au+Au$ collisions   
by the STAR Collaboration at RHIC using the $z$-scaling approach are presented. 
Indication on self-similarity of fractal structure of nuclei 
and fragmentation processes with  $K^0_S$ probe 
is demonstrated. The energy loss as a function of the collision energy, centrality and transverse
momentum of the inclusive strange meson is estimated.
}


\section{Introduction}
\label{sec:intro}

The idea of self-similarity of hadron interactions is a fruitful concept 
to study collective phenomena in hadron matter.
Important manifestation of such a concept is existence of scaling itself.  
Some of the scaling features constitute pillars of modern critical phenomena. 
Other category of scaling laws (self-similarity in point explosion, laminar and turbulent 
fluid flow, super-fluidity far from phase boundary and critical point, etc.) 
reflects features not related to phase transitions.
The notions "scaling" and "universality" have special importance in critical phenomena. 
The universality hypothesis reduces the great variety of critical phenomena to a small number 
of equivalence classes, the so-called "universality classes", which depend only 
on few parameters (critical exponents). 
The universality has its origin in the long range character of interactions. 
Close to the transition point, the behavior of the cooperative phenomena 
becomes independent of the microscopic details of the considered system. 
The fundamental parameters determining the universality class are 
the symmetry of the order parameter and the dimensionality of space.

\section{General concept of $z$-scaling}

The concept of the $z$-scaling is based on the fundamental principles of self-similarity, 
locality, and fractality of particle production 
in $p+p$ and $A+A$ interactions at a constituent level \cite{1,2,3,4}. 
At sufficiently high energies, the collision of hadrons or nuclei 
is considered as an ensemble of individual interactions of their constituents.
Structures of the colliding objects are characterized by
fractal dimension $\delta_1$ and $\delta_2$.
The interacting constituents carry the fractions
$x_1$ and $x_2$ of the momenta $P_1$ and $P_2$ 
of the colliding hadrons (or nuclei).
The fragmentation of the scattered constituent and fragmentation of the produced recoil is described
by fractal dimensions $\epsilon_a$ and $\epsilon_b$, 
and by the momentum fractions $y_a$ and $y_b$, respectively.
The gross features of the momentum distributions of produced  inclusive particles 
can be described in terms of the kinematic
characteristics of the corresponding  sub-processes. 
The sub-process is considered as a binary collision
of constituents with masses $x_{1}M_{1}$ and $x_{2}M_{2}$, which results in
the scattered and recoil object in the final state with masses $m_a/y_a$ and 
$M_X=x_{1}M_{1}+x_{2}M_{2} + m_b/y_b$, respectively.
The associate production of the particle with mass~$m_b$ ensures conservation
of additive quantum numbers.
The momentum conservation in the sub-process
is connected with the recoil mass as follows $(x_1P_1\!+\!x_2P_2\!-\!p/y_a)^2\!=\!M_X^2$.
A function of the momentum fractions,   
$\Omega =
(1-x_1)^{\delta_1}(1-x_2)^{\delta_2}(1-y_a)^{\epsilon_a}(1-y_b)^{\epsilon_b}$, 
is introduced, which is proportional to
the number of all such constituent configurations, which contain the configuration defined 
by the fractions $x_1$, $x_2$, $y_a$ and $y_b$.
It plays the role of a relative volume which occupy these configurations
in space of the momentum fractions.
The self-similarity variable $z$ and the scaling function $\psi(z)$ are expressed via model parameters,
momentum fractions, multiplicity density, total inelastic and inclusive cross section of produced hadron, 
and the corresponding Jacobian J  \cite{1,2,3,4}: 
$z =z_0 \Omega^{-1}$,  $\psi(z) = { \pi }/{ [(dN/d\eta)\sigma_{inel}} J] \cdot E{{d^3\sigma}/ {dp^3}}$.

 \section{Self-similarity of $K_S^0$ production in $Au+Au$ collisions}
 
\begin{figure*}[h]
\begin{center}
\hspace{-1mm}
{\includegraphics[width=43mm,height=43mm]{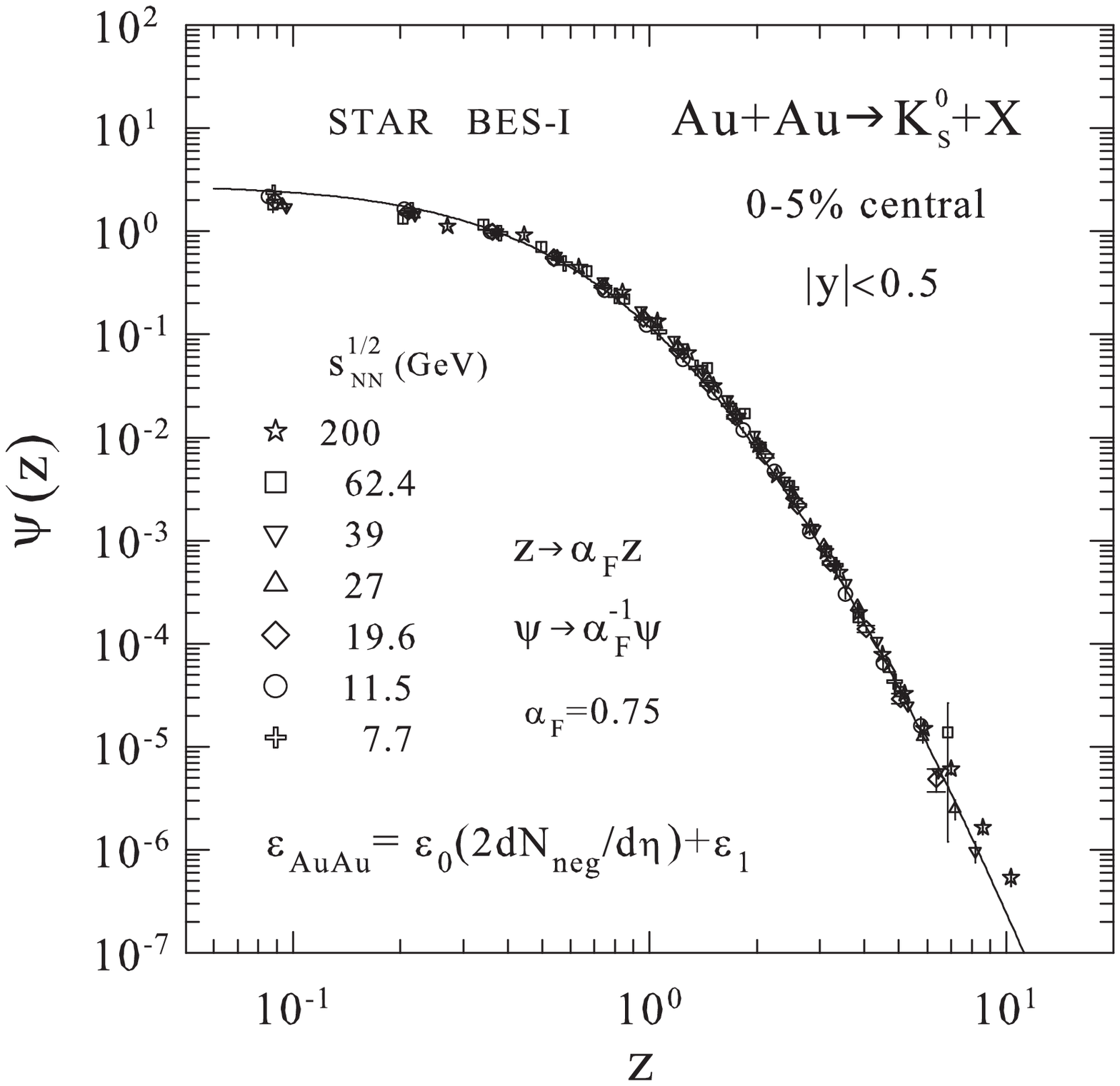}}
\hspace{11mm} 
{\includegraphics[width=41mm,height=43mm]{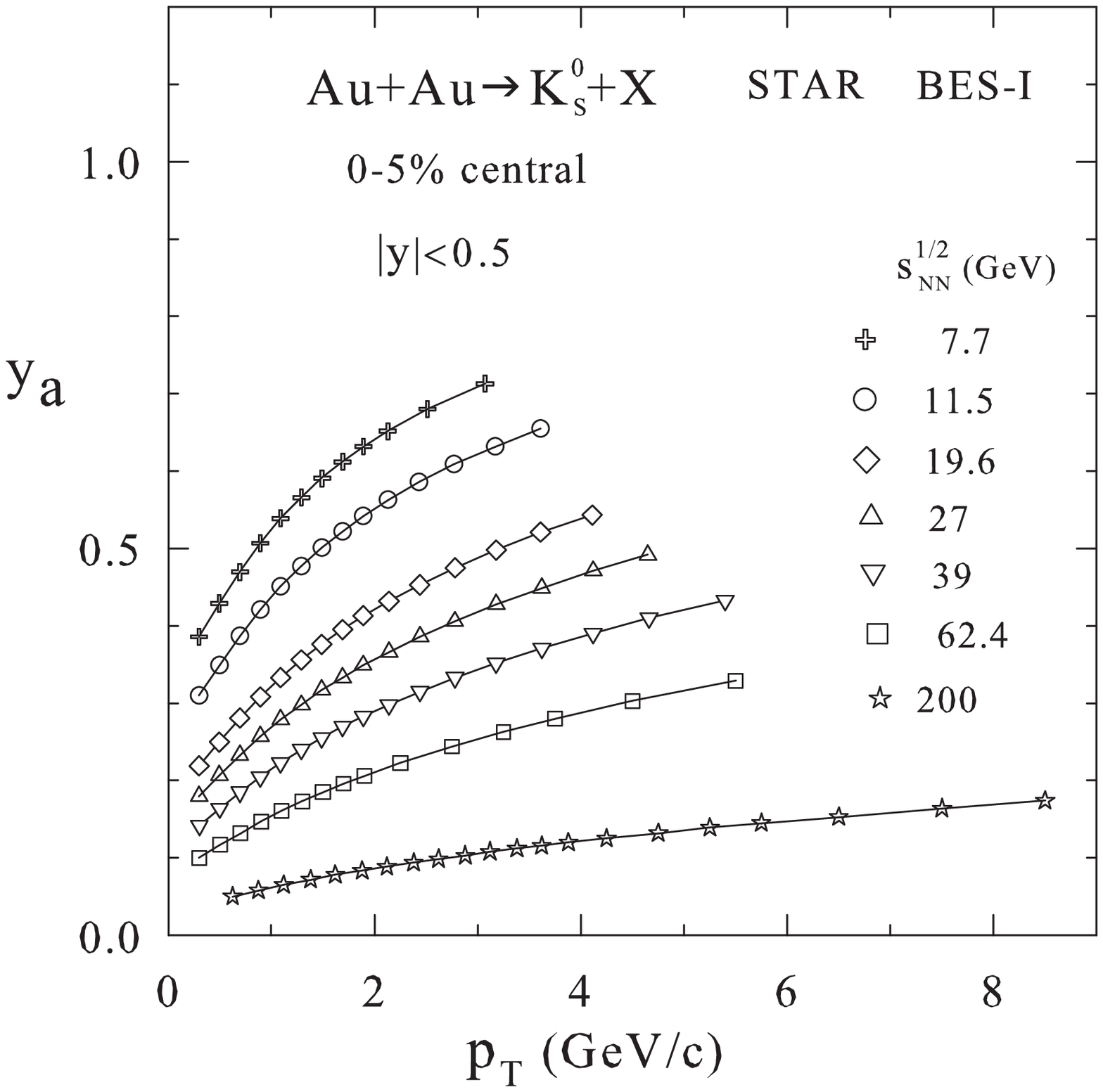}}
\vskip 3mm
\hspace{0.2cm}   a) \hspace*{47mm} b) 
\vskip -2mm
\caption{\label{fig:1}
The scaling function $\psi(z)$ (a) and momentum fraction $y_a$  (b)  
for $K^0_S$ mesons produced 
in $(0-5)\%$ central $Au+Au$ collisions 
at $\sqrt {s_{NN}}=7.7, 11.5, 19.6, 27, 39, 62.4, 200$~GeV 
in the rapidity interval $|y|<0.5$. 
The symbols correspond to data
obtained by the STAR Collaboration at RHIC.
The solid line is a reference curve for $p+p$ interactions.
}
\end{center}
\end{figure*}

 Figure \ref{fig:1}(a) shows the dependence of the scaling function $\psi(z)$ 
on the variable $z$ for $K^0_S$ mesons  
produced in the $(0-5)\%$ central $Au+Au$ collisions for different energies.
The symbols correspond to $p_T$-distributions 
measured in the $Au\!+\!Au$ system
and the solid line is the $z$-scaling curve for $p+p$ interactions.
Reasonable coincidence of the symbols and the solid curve in the interval $z=0.1-7$ was found
provided specific dependencies of model parameters on energy and multiplicity.
There are no irregularities in the behavior of $\psi(z)$ 
over a wide range of collision energy $\sqrt {s_{NN}}=$ 7.7 - 200~GeV.
An indication on a flattening of the scaling function 
at low $z$ and a power-law at high $z$ is clearly observed.
Similar results were obtained for other collision centralities.
This result is interpreted as a  self-similar modification 
of the constituent sub-processes by the created medium.

The energy loss is characterized by the fraction $y_a$ in the $z$-scaling scheme.
The quantity is sensitive characteristic of the nuclear medium.  
Figure \ref{fig:1}(b) shows the dependence of the momentum fraction $y_a$ 
on the transverse momentum $p_T$ of $K^0_S$ mesons produced 
in the most central $(0-5)\%$ $Au+Au$ collisions at different energies. 
A monotonic growth of $y_a$ with $p_T$ is found for all energies 
and all collision centralities.
This means that the relative energy dissipation $\Delta E_q/E_q=(1-y_a)$ associated with 
a high-$p_T$ particle is smaller than for processes with lower transverse momenta.
The energy loss becomes larger as the collision energy increases.
It was found that 
the energy loss is greater in the most central $Au+Au$ collisions 
than in the peripheral ones.
The energy losses alone, which have been estimated for 
the production of $K_S^0$ mesons
in $Au+Au$ collisions over a wide range of collisions energy, centrality and $p_T$, 
show no sign of a phase transition.

\section{Conclusion}
We analyzed data on transverse momentum spectra of $K_S^0$  mesons 
measured in $Au+Au$ collisions at mid-rapidity 
by the STAR Collaboration in BES-I program at RHIC
in the $z$-scaling approach.
Self-similarity of $K_S^0$-meson production in the gold-gold collisions  
over a wide kinematic and centrality range was found.    
Constituent energy loss as a function of the collision energy, 
centrality and transverse momentum of $K_S^0$ mesons  was estimated.
The method of  analysis is extended  for systematic description 
of $A+A$ collisions with production of identified hadrons.     

\section*{Acknowledgements}
This work was partially supported by Project funded by
the MEYS of the Czech Republic under the contract LTT18021.



\nolinenumbers

\end{document}